# XNOR-VSH: A Valley-Spin Hall Effect-based Compact and Energy-Efficient Synaptic Crossbar Array for Binary Neural Networks

Karam Cho and Sumeet Kumar Gupta, *Senior Member, IEEE*

**Binary neural networks (BNNs) have shown an immense promise for resource-constrained edge artificial intelligence (AI) platforms as their binarized weights and inputs can significantly reduce the compute, storage and communication costs. Several works have explored XNOR-based BNNs using SRAMs and nonvolatile memories (NVMs). However, these designs typically need two bit-cells to encode signed weights leading to an area overhead. In this paper, we address this issue by proposing a compact and low power in-memory computing (IMC) of XNOR-based dot products featuring signed weight encoding in a single bit-cell. Our approach utilizes valley-spin Hall (VSH) effect in monolayer tungsten di-selenide to design an XNOR bit-cell (named 'XNOR-VSH') with differential storage and access-transistor-less topology. We co-optimize the proposed VSH device and a memory array to enable robust in-memory dot product computations between signed binary inputs and signed binary weights with sense margin (SM) > 1 $\mu$A. Our results show that the proposed XNOR-VSH array achieves 4.8% ~ 9.0% and 37% ~ 63% lower IMC latency and energy, respectively, with 49% ~ 64% smaller area compared to spin-transfer-torque (STT)-MRAM and spin-orbit-torque (SOT)-MRAM based XNOR-arrays.**

## I. INTRODUCTION

Deep neural networks (DNNs) have realized remarkable advances for artificial intelligence (AI) workloads achieving super-human accuracies in several tasks. However, this comes at enormous storage, computation, communication costs as traditional solutions based on the standard architectures (e.g., graphics processing unit (GPU) and tensor processing unit (TPU)) entail a large number of power-hungry and performance-limiting processor-memory transactions. Therefore, to efficiently handle the data-intensive workloads in DNNs, in-memory computing (IMC) architectures (where computing operations are performed inside a memory macro) have been extensively investigated [1]–[4]. The IMC reduces frequent data movement between memory and processor yielding much lower communication costs.

However, for highly energy-constrained edge AI platforms IMC alone may not be sufficient to meet the energy efficiency targets. To manage the power consumption of edge devices, quantization of the inputs and weights is a common technique [4], which reduces the storage and IMC energy costs drastically. In fact, input/weight quantization all the way to binary levels has been shown to significantly boost the IMC energy efficiency for DNNs. The simplest binary quantization scheme involves encoding the weights and inputs as 0s and 1s (unsigned binary); however, such approaches suffer from large accuracy loss. To bring the accuracies to acceptable levels (while still reaping the energy benefits of heavy quantization), signed binary neural networks (BNNs) have emerged where weights and inputs are binarized to +1 and -1 [5]. The IMC of scalar product of weights and inputs corresponds to an XNOR operation; hence, such designs are referred to as XNOR-BNNs.

To implement XNOR-based signed BNNs, there are two common techniques. One is to encode signed binary weights into two bit-cells in the memory array to achieve complementary weight encoding. Such an approach has been explored in the context of SRAM [6] and emerging nonvolatile memories (NVMs) including resistive random-access memory (XNOR-RRAM [7]) and magnetic tunnel junction (MTJ)-based magnetic random-access memories (MRAMs). The MTJ-based designs exploit spin-transfer-torque (XNOR-STT [8], [9]) or spin-orbit-torque (XNOR-SOT [10]). However, the main bottleneck of this approach is the area overhead due to the need for two bit-cells per a set of complementary weights. To mitigate this issue, the second technique utilizes a single bit-cell to store weights in the forms of 0s and 1s and performs IMC in the unsigned regime. However, it requires a transformation of the outputs from the unsigned regime to the signed regime, which entails pre-processing of inputs/weights and post-processing of outputs leading to peripheral circuit overheads [11]–[13].

In this paper, we address the limitations of the existing solutions for XNOR-BNNs by utilizing valley-spin Hall (VSH) effect in monolayer tungsten di-selenide to perform XNOR-based IMC (named 'XNOR-VSH'). Our design features 1) encoding signed binary weights in the array (thus, avoiding the post-processing overheads associated with transformations between signed and unsigned regimes), and 2) storing complementary bits in a *single* device (thus, averting the need for two bit-cells for signed weight encoding). Moreover, the proposed design leverages the integrated back-gate in VSH device (which we explored in our earlier works [14]) to achieve

This work was supported by SRC/NIST-funded NEWLIMITS Center (Award number 70NANB17H041) and SRC/DARPA-funded CoCoSys Center.

Karam Cho and Sumeet Kumar Gupta are with Elmore Family School of Electrical and Computer Engineering, Purdue University, West Lafayette, IN 47907, USA (e-mail: cho346@purdue.edu; guptask@purdue.edu).

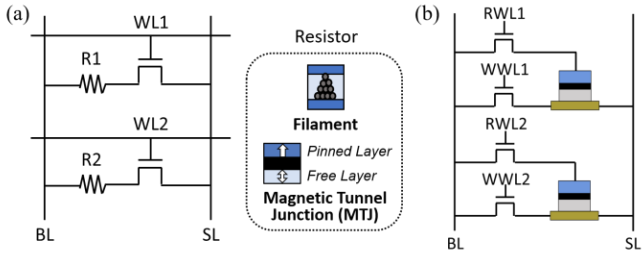

Fig. 1. State-of-the-art XNOR: (a) 2T-2R XNOR cell based on RRAM (R as filament) [7] or STT-MRAM (R as MTJ) [8], [9] and (b) SOT-MRAM-based XNOR cell with 4T-2R configuration [10].

access-transistor-less design, leading to further area savings. We will show later that the compactness in the proposed design enabled by complementary bit encoding in a single device along with the integrated back-gate [14] translates to a significant reduction in IMC energy and latency compared to the previous spin-based XNOR-IMC alternatives. The key contributions of this paper are as follows:

- We propose an XNOR bit-cell utilizing the VSH effect (XNOR-VSH) where complementary signed weights (+1/-1) are stored in a *single* device enabling an area-efficient layout compared to the state-of-the-art 2T-2R XNOR designs.
- We implement an array based on the proposed XNOR-VSH bit-cell that can perform IMC of dot products of signed binary weights and inputs satisfying sense margin (SM) > 1 $\mu$A.
- We evaluate the benefits and trade-offs of XNOR-VSH vis-à-vis XNOR-STT and XNOR-SOT, showing significant area, energy and latency benefits of the proposed technique at the cost of lower SM.

## II. BACKGROUND

### A. State-of-the-art XNOR-IMC based on Nonvolatile Memories

Figure 1 shows the state-of-the-art XNOR cell designs based on emerging NVMs. Fig. 1(a) depicts 2T-2R XNOR cell where R1 and R2 describe resistive components, which can be either memristors or MTJs, representing XNOR-RRAM [7] or XNOR-STT [8], [9], respectively. In these designs, synaptic weights are encoded in the resistive components that usually have two resistance levels depending on their configuration. For XNOR-RRAM, the memristor becomes conducting (e.g., a filament is formed between two electrodes) when a set voltage across it is greater than the threshold voltage, leading to low resistance ($R_L$). When reset (e.g. the filament is broken), the memristor loses its conducting characteristics and exhibits high resistance ($R_H$). By utilizing two 1T-1R RRAM bit-cells, complementary synaptic weights are stored in the XNOR cell. Weight = +1 (-1) corresponds to $R_L$ ($R_H$) stored in one bit-cell and $R_H$ ($R_L$) in the other. The input is encoded by driving the two word-lines (WLs) of the XNOR cell to $V_{DD}$ or 0. Input = +1 (-1) corresponds $V_{DD}$ (0) on one WL and 0 ($V_{DD}$) on the other. The combination of input vectors and stored weights returns the XNOR output on the bit-line (BL) or sense-line (SL). This XNOR output current from multiple XNOR-cells is summed on the BL/SL to obtain the dot product of the inputs and weights in the signed binary regime.

The MTJs in XNOR-STT play a similar role in storing the weights. When the write current flowing through an MTJ is large enough to switch the magnetization of the free layer (FL) aligning it with that of the pinned layer (PL), MTJ is in parallel (P) state with $R_L$ as its resistance. If the FL is set to have the opposite magnetization as that of the PL (by flowing the write current in the opposite direction to the former case), it holds anti-parallel (AP) state with $R_H$ as its resistance. Thus, XNOR-STT also requires two bit-cells to store complementary weights and sense the XNOR output in the same way as XNOR-RRAM. In case of XNOR-SOT (see Fig. 1(b)) [10], encoding the weights is achieved by flowing write currents along the heavy metals (HMs) such that in-plane spin polarizations are induced at the interface between the HMs and FLs of MTJs, exerting SOT on the FLs. XNOR-SOT is more write-efficient than XNOR-STT and allows independent optimization of write and IMC, but requires additional access transistors. This results in a 4T-2R configuration and thus, larger cell area. The IMC in XNOR-SOT is performed in a similar manner as XNOR-RRAM by flowing the current in the read path through the MTJs.

However, such two-bit-cell-based designs can lead to non-trivial area overheads. To relieve the issue, relevant works have been conceived to utilize a single bit-cell to store weights in the forms of 0s and 1s and perform IMC in the unsigned regime [11]–[13]. In the NAND-net [11], XNOR operations are replaced by NAND operations, requiring only one bit-cell per weight and simplifying the IMC of dot products. In [12], [13], the signed inputs are replaced by the unsigned inputs to improve the area and energy efficiencies. However, such techniques require transformations of the outputs from the unsigned regime to the signed regime, which involve pre-processing of inputs/weights and post-processing of outputs leading to the need for extra peripheral circuits.

Although the state-of-the-art XNOR designs exhibit higher density than CMOS-based designs such as XNOR-SRAM [6], they either suffer from the area overheads due to the necessity of multiple transistors to enable XNOR-IMC or possess circuit complexity arising from the transformations of inputs/outputs between unsigned and signed regimes. These limit the scalability and the IMC energy efficiency. In this paper, we propose VSH-based XNOR-IMC, which avoids the overheads of both the techniques by enabling signed binary weight encoding in a single device and designing compute-enabled array for XNOR-IMC without the overheads of transformations between signed and unsigned regimes.

### B. Valley-Spin Hall (VSH) Effect in Monolayer WSe2

Monolayer transition metal dichalcogenides (TMDs) such as tungsten di-selenide ($WSe_2$) exhibit intriguing spintronic features promising low power applications [14]–[17]. Due to the unique physics of monolayer TMDs, an external electric field (or charge current, $I_C$) generates transverse spin current ($I_S$) by separating carriers with opposite out-of-plane (i.e., up or down) spins in divergent directions, which is known as valley-spin Hall (VSH) effect [14]–[18]. Given the generation of out-of-plane spins, perpendicular magnetic anisotropy (PMA) magnets can be coupled with $WSe_2$ to switch their magnetization utilizing VSH-effect-driven SOT (VSH-SOT) without any external magnetic field. As the PMA magnets are known to be more energy-efficient in switching than in-plane magnetic anisotropy (IMA) magnets, VSH promises to reduce write energy compared to IMA-based devices (e.g., SOT-MRAM). Therefore, energy-efficient nonvolatile memories based on the VSH effect (VSH-MRAM; see the inset of Fig. 2) were previously proposed in [14] by utilizing the WSe$_2$ spin generator, featuring an integrated back-gate, and coupling it with the PMA MTJs. These designs, targeting memory operations and IMC of Boolean and simple arithmetic functions (for general purpose computing) showed higher energy-efficiency along with access-transistor-less compact layout due to the integrated back-gate. In this work, we utilize the VSH effect in monolayer WSe$_2$ to design an array that can compute XNOR-based dot products with signed binary weights and inputs (targeting BNNs).

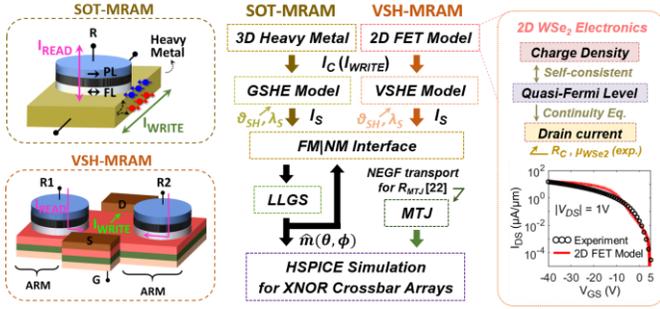

Fig. 2. Simulation framework for evaluation and comparison of the proposed XNOR-VSH with STT- and SOT-based XNOR designs. Details of the 2D FET model is in [18].

TABLE I
PARAMETERS FOR HSPICE SIMULATION

| Parameter | STT (PMA) | SOT (IMA) | VSH (PMA) |
|---|---|---|---|
| MTJ Size, $L \times W$ [nm$^2$] | 30×30 | 25×75 | 30×30 |
| MTJ Thickness, $T_M$ [nm] | 1.3 | 1.7 | 1.3 |
| MgO Thickness [nm] | 1.1 | 1.3 | 1.3 |
| Saturation Magnetization, $M_S$ [emu/cm$^3$] | 1257.3 [31] | 1257.3 [31] | 1257.3 [31] |
| Uniaxial Anisotropy Density, $K_U$ [erg/cm$^3$] | 2.3×10$^6$ [32] | 8.3×10$^5$ [32] | 2.3×10$^6$ [32] |
| Energy Barrier, $E_B$ [$k_B$T] | ~50 | ~50 | ~50 |
| Damping Coefficient, $\alpha$ | 0.008 | 0.007 | 0.008 |
| Gyromagnetic Ratio, $\gamma$ [MHz/Oe] | 17.6 | 17.6 | 17.6 |

### C. Simulation Framework

We establish a simulation framework (Fig. 2) to evaluate the proposed VSH-based XNOR-IMC (XNOR-VSH), and to perform a comparison with the STT- and SOT-MRAM-based XNOR designs (XNOR-STT and XNOR-SOT, respectively). Note, STT- and SOT-MRAMs are also optimized to implement XNOR dot products for fair comparison with the proposed design, the details of which will be discussed in section IV.

For XNOR-VSH, we utilize the model developed previously by us and described in detail in [14]. Here, we provide a brief overview of the model. First, a 2D FET model is developed based on the approach in [18] to self-consistently capture 2D electrostatics and charge transport in 2D WSe$_2$ channel of the VSH device. The drain current (or charge current, $I_C$) is calculated as a function of the gate ($V_{GS}$) and drain voltages ($V_{DS}$) with mobility of WSe$_2$ ($\mu_{WSe2}$) [18] and contact resistance ($R_C$) at S/D side [19] calibrated with the experiments. During *Write*, spin current, $I_S$ is obtained from $I_C$ considering valley-spin Hall angle ($\theta_{SH}$) and spin diffusion length ($\lambda_S$) of WSe$_2$ layer, extracted from the experiments [16], [17]. Generated $I_S$ is provided to ferromagnet | non-magnet (FM|NM) interface model [20] and Landau-Lifshitz-Gilbert-Slonczewski (LLGS) equation [20] to capture the interface spin scattering and resultant switching dynamics of the PMA magnets, which serve as the FLs of MTJs sitting on the WSe$_2$ layer. The MTJ resistance which is utilized for IMC operation of the proposed design is obtained from the Non-Equilibrium Green Function (NEGF) equations, as detailed in [21]. Finally, HSPICE simulation evaluates energy and delay of the proposed XNOR-VSH array. Note, during *Read* and XNOR-IMC, a distributed resistance network is used to model the current paths via WSe$_2$ channel, which captures the unique geometry of the VSH device as detailed in [14]. The resistance is obtained from 2D FET model similar to [14] as a function of $V_{GS}$ and $V_{DS}$ capturing Schottky contact resistance between MTJs and WSe$_2$.

For comparison, XNOR-STT and XNOR-SOT are implemented in HSPICE based on the cell designs in Fig. 1. For XNOR-STT, the LLGS equation [20] is used to capture the effect of spin torque in MTJs. XNOR-SOT follows the same simulation framework as XNOR-VSH except that the $I_C$ along the HM (here, tungsten W [22]) is calculated, provided to a GSHE model, and converted into $I_S$ via $\theta_{SH}$ of W, which switches IMA magnets (FLs of MTJs in XNOR-SOT). The MTJ resistance model similar to XNOR-VSH is utilized in the IMC operation of XNOR-STT/SOT. For both designs, 7-nm n-FinFETs [low power predictive technology model (PTM) model] are used for access transistors. Table I shows the parameters used in HSPICE simulation. Here, STT- and VSH-based XNORs utilize PMA magnets while XNOR-SOT is designed with IMA magnet. All magnets are optimized to have energy barrier ($E_B$) of ~50 $k_BT$ (where $k_BT$ is the thermal energy).

## III. PROPOSED XNOR-VSH

### A. XNOR-VSH Bit-cell

We design the XNOR-VSH bit-cell by utilizing the differential storage capability of VSH-MRAM explored previously in [14]. However, unlike the work in [14] (which focused on utilizing the differential functionality for storage and IMC of Boolean functions/addition of in-memory operands for general purpose computing), here, we harness the VSH effect for implementing XNOR-IMC for BNNs. Therefore, the routing and biasing of various lines in the proposed bit-cell and the optimization strategies are different from [14] and will be described shortly.

The XNOR-VSH bit-cell includes a p-type spin generator, which has monolayer WSe$_2$ channel with an integrated back-gate (see inset in Fig. 2). The channel is extended beyond the area between source and drain, forming two arms. The back-gate controls the flow of charge and spin currents ($I_C$ and $I_S$) between S and D, and in the arms, respectively. As $I_C$ (or $I_{WRITE}$) flows from S to D, the carriers of opposite spins ($I_{S\uparrow}$ or $I_{S\downarrow}$) diverge due to VSH effect, leading to the flow of $I_S$ in the arms transverse to $I_C$. The induced $I_S$ interacts with PMA magnets, which serve as FLs in MTJs located on each arm, storing complementary bits (as discussed shortly). For reading the bits stored, read paths are formed along the MTJs and WSe$_2$ channel via the read nodes (R1 and R2) formed on top of the MTJs.

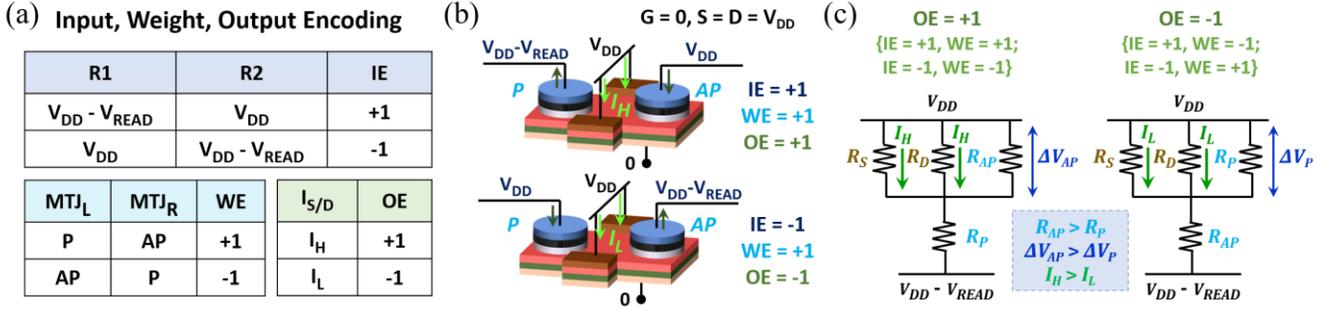

Fig. 3. (a) Truth table of the proposed XNOR-VSH binary encoding of input (IE), weight (WE), and output (OE). (b) Illustration for examples of OE = +1 (upper) and OE = -1 (lower). (c) Equivalent circuit for sensing of OE = +1 (left) and OE = -1 (right).

In our proposed XNOR-VSH, both the weight *W* and input *IN* are binarized to +1 or -1, so that their scalar multiplication (i.e., output *OUT*) can represent the bitwise XNOR operation. First, *W* encoding (WE) is achieved following the *Write* operation. Complementary bias voltages (i.e., $V_{DD}$ or 0) are applied at S and D, following which the p-type spin generator is turned ON (G driven to 0). For example, when S is driven to $V_{DD}$ and D to 0, $I_{WRITE}$ flows from S to D inducing $I_{S\downarrow}$ and $I_{S\uparrow}$ in the left and right arms, respectively. This switches the left FL downward and the right FL upward. Hence, the MTJs on the left (MTJ$_L$) and right (MTJ$_R$) are set to AP and P states, respectively. Note, the PLs of MTJs are fixed to +z. We define this MTJ configuration as *W* = -1 (see Fig. 3). The opposite WE (*W* = +1) is achieved by reversing the direction of $I_{WRITE}$ with the biasing of S at 0 and D at $V_{DD}$, which leads to $I_{S\uparrow}$ and $I_{S\downarrow}$ in the left and right arms, respectively, storing P in MTJ$_L$ and AP in MTJ$_R$. It is important to note that the complementary WE is implemented in a *single* XNOR-VSH bit-cell. This, along with the integrated back-gate, yields a compact layout compared to the existing XNOR designs where, in general two bit-cells in the 2T-2R configuration are required to realize the complementary WE (see Fig. 1) [7]–[10]. As a result, the proposed XNOR-VSH design yields significant area savings and the resultant IMC energy-efficiency, which will be discussed in section IV.

Next, we define the *IN* encoding (IE) and explain the implementation of XNOR-functionality in the bit-cell in situ. *IN* is '+1' when the read node R1 is driven to '$V_{DD} - V_{READ}$' and R2 to '$V_{DD}$.' The opposite biasing (i.e., R1 and R2 driven to '$V_{DD}$' and '$V_{DD} - V_{READ}$,' respectively) represents *IN* of '-1.' To compute the XNOR-based scalar product of *IN* and *W*, we apply $V_{DD}$ at both S and D, drive R1 and R2 to either '$V_{DD} - V_{READ}$' or '$V_{DD}$' (depending on the *IN* value) and assert the back-gate (G = 0). Note, with this biasing scheme, the voltage drop across MTJs is limited to $V_{READ}$, which is small enough so as not to disturb the stored *W* during the compute. The XNOR-based scalar product between *IN* and *W* is obtained by sensing the current flowing through S (or D). The proposed biasing leads to current paths along S, D, and the two MTJs (and their respective read nodes R1 and R2), which is explained as follows. A node connected to '$V_{DD} - V_{READ}$' acts like a sink for current while the other three nodes inject the current in the WSe$_2$ channel. Recall, this current flow is modeled using the distributed non-linear $V_{GS}/V_{DS}$-dependent resistance network (as mentioned in section II), and our results are based on self-consistent simulations accounting for various physical effects. However, to explain the concept, let us simplify the discussion by considering an equivalent circuit shown in Fig. 3(c). Note, the current sensed corresponds to that flowing through $R_S$. When *IN* = +1 and *W* = +1, S, D, and node R2 (connected to AP-MTJ) are driven to $V_{DD}$ while the node R1 (connected to P-MTJ) is driven to $V_{DD} - V_{READ}$, making R1 the sink for current. In this case, the voltage drop across $R_S$ is determined by the voltage division between $R_S // R_D // R_{AP}$ and $R_P$. Since $R_P < R_{AP}$, the voltage drop across $R_S$ (annotated as $\Delta V_{AP}$ in Fig. 3(c)) is large, leading to high sensing current through $R_S$ ($I_H$). This corresponds to *OUT* = +1. A similar configuration is obtained when *IN* = -1 and *W* = -1, with the only exception that now, node R2 acts like a sink and R1 as a source for the current. Let us now consider the two cases: (i) *IN* = +1 and *W* = -1 and (ii) *IN* = -1 and *W* = +1. In both the scenarios, the read node connected to $R_P$ acts as a current source while that connected to $R_{AP}$ serves the role of current sink. Now, the voltage division action occurs between $R_S // R_D // R_P$ and $R_{AP}$, leading to lower voltage across $R_S$ ($\Delta V_P$) and low sensing current ($I_L$), which corresponds to *OUT* = -1. Thus, when *IN* and *W* are the same, *OUT* = +1 is obtained, while when they are different (i.e., opposite in sign), *OUT* = -1 is computed, thus implementing the XNOR functionality. Note, for robust operation, $I_H$ needs to be sufficiently larger than $I_L$, which we ensure by design, as discussed subsequently.

### B. XNOR-VSH Array

Utilizing the proposed XNOR-VSH bit-cell, we design an array with 64 rows and 64 columns (see Fig. 4). The back-gate, S, and D of the bit-cell are connected to WL, BL, and BL-bar (BLB), respectively. The read nodes (R1 of MTJ$_L$ and R2 of MTJ$_R$) are connected to read-word-line A (RWL$_A$) and read-word-line B (RWL$_B$). RWLs are associated with the input activation and are routed along the row. BL/BLBs run along the column and are used to program the weight during write. WLs are routed along the row and asserted during both write and read to make the WSe$_2$ channel conducting. The IMC output current is sensed at BL. In this work, driver resistance of 0.5 kΩ and sensing resistance of 0.1 kΩ are considered, following the work in [23].

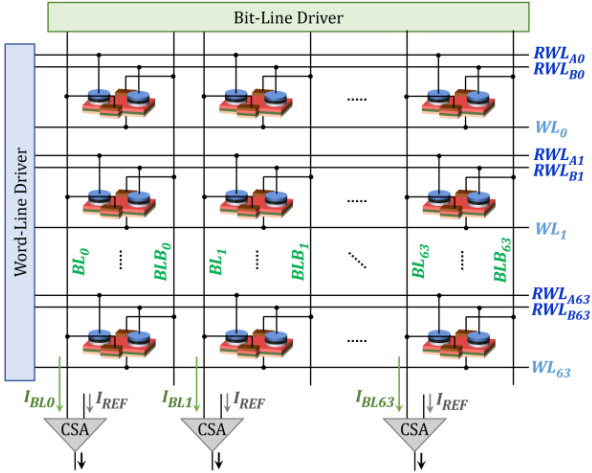

Fig. 4. An XNOR-VSH array with 64 rows and 64 columns. Each column performs MAC operation.

TABLE II
OPERATING BIAS FOR XNOR-VSH ARRAY

| | | Write | Read/Compute |
|---|---|---|---|
| Accessed | WL | 0 | 0 |
| | RWL$_A$ | V$_{DD}$ (floating) | V$_{DD}$/V$_{DD}$-V$_{READ}$ |
| | RWL$_B$ | V$_{DD}$ (floating) | V$_{DD}$-V$_{READ}$/V$_{READ}$ |
| | BL | V$_{DD}$/0 | V$_{DD}$ |
| | BLB | 0/V$_{DD}$ | V$_{DD}$ |

* During **Hold**, all lines are pre-charged to V$_{DD}$.

### 1) Write Operation

In BNNs, the trained binary weights are loaded into the array during the write/program operation. For this, BL and BLB of the cell are driven to $V_{DD}$ (0.9 V) or 0 V according to the bit-information which needs to be stored. Then, we apply 0 V to the active-low WL of the selected bit-cell (recall that the proposed XNOR device is p-type). The BL/BLB biasing determines the direction of $I_{WRITE}$ in the accessed bit-cell, and writes P/AP-states into MTJs in the arms, as described before. For example, when we write $W$ = -1 into the bit-cell, BL and BLB are driven to $V_{DD}$ and 0 V, respectively. To encode $W$ = +1 in the bit-cell, BL and BLB are driven to 0 V and $V_{DD}$. As discussed before, the complementary weight encoding is feasible within one bit-cell in the proposed XNOR-VSH design. Here, RWL$_A$ and RWL$_B$ are kept pre-charged (and floating) at $V_{DD}$ to keep most of the $I_{WRITE}$ between S and D only. For unaccessed cells, all lines are driven to $V_{DD}$.

### 2) XNOR-IMC and Read Operations

In the proposed XNOR-VSH array, the dot product output is read by sensing the currents flowing through BL ($I_{BL}$). As noted before, the signed weights are encoded in the MTJs of XNOR-VSH bit-cells, and the signed inputs are applied via corresponding RWL$_A$ and RWL$_B$. Also, recall from the previous sub-section that the proposed bit-cell and biasing scheme implement the XNOR functionality, which corresponds to the scalar product of the input and weight in the signed binary regime.

To compute the dot product of input vector and weight matrix, we activate multiple WLs ($N$) and sense the accumulated $I_{BL}$ realizing the MAC operation within a column. Similar to [7], the analog summation of $I_{BL}$ can be expressed using the number of bit-cells with $I_H$ ($a$) and $I_L$ ($N$ - $a$):

$$I_{BL} = aI_H + (N - a)I_L \quad (1)$$

Now, by replacing $I_H$ and $I_L$ with the corresponding bit values of +1 and -1, respectively, the MAC output ($OUT_M$) of the column is obtained as:

$$OUT_M = a(+1) + (N - a)(-1) = 2a - N \quad (2)$$

The $I_{BL}$ of the column is digitized using an analog-to-digital converter (ADC) [24]. From this, the corresponding value of $a$ can be obtained following (1), and $OUT_M$ can be deduced from (2). These operations correspond to the bit-counting technique proposed in [5]. The illustration in Fig. 5(a) describes the IMC operation in the proposed XNOR-VSH using an example with 8 bits, which can be implemented by simultaneously activating 8 rows in the XNOR-VSH array. For this example, $IN$ and $W$ vectors are {-1, +1, -1, -1, +1, -1, -1, +1} and {+1, +1, -1, -1, -1, -1, +1, +1}, respectively. Each bit-cell produces currents according to their $IN$ and $W$ that are {$I_L$, $I_H$, $I_H$, $I_H$, $I_L$, $I_H$, $I_L$, $I_H$}, representing $OUT_M$ vector of {-1, +1, +1, +1, -1, +1, -1, +1}. From the accumulated $I_{BL}$, we can deduce $a$ to be 5 (from (1)). Then, we calculate the output of '$2a - N = 10 - 8 = +2$' (from (2)) which matches the result of regular MAC operation. Therefore, $I_{BL}$ of a single column can be translated into $a$ using the ADC, and finally $OUT_M$ in the proposed XNOR-VSH can be achieved using digital peripheral circuits implementing the shift operation and subtraction.

It may be noted that reading the stored weight in the array is a special case of the XNOR-IMC (i.e., $IN$ = +1 with only one row activated). Therefore, we can apply the biases corresponding to $IN$ = +1 (as described above) and sense $I_{BL}$ to perform the read operation.

### 3) IMC Robustness

For efficient data processing in a BNN array, performing XNOR-IMC in multiple columns is a typical approach which offers the benefits of parallelism. In our proposed XNOR-VSH array, we follow the same methodology by simultaneously asserting all the columns of the array and reading their MAC outputs in one shot. However, this leads to interactions between different cells in an array due to sneak current paths (as also in several other solutions based on crossbar arrays). In other words, the cross-point connections between RWLs and BLs can make the output current to be data-dependent as the equivalent resistance of the array seen by a column can be different for different weight combinations. Hence, we need to ensure by design that the output of any column is not sullied by the interaction with other cells in the array.

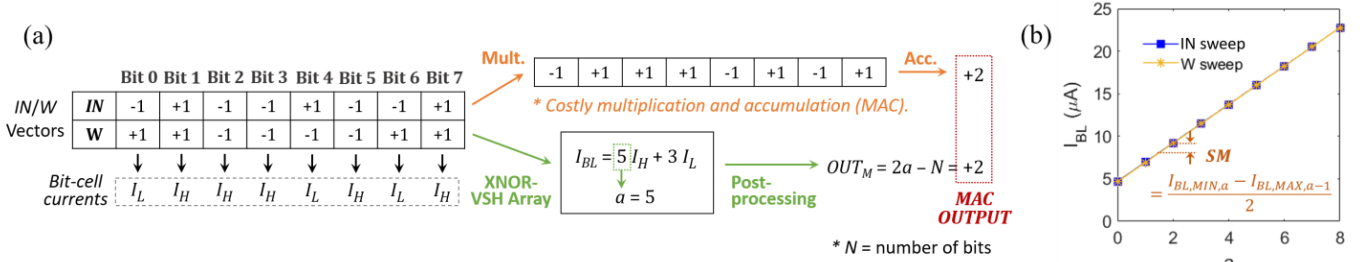

Fig. 5. (a) Illustration of bitwise XNOR operation for *N*-bit *IN* and *W* vectors (here, *N* = 8) in the proposed XNOR-VSH array replacing the costly MAC operation. (b) $I_{BL}$ of column 0 with 8 rows ($N$ = 8) and 64 columns asserted in the XNOR-VSH array (array size = 64*64). Columns 1:63 hold equal distributions of '+1' and '-1' weights.

To verify the compute functionality in the proposed XNOR-VSH array, we assert 8 rows and all 64 columns (i.e., maximum column-wise parallelism), and sense the $I_{BL}$ of all columns. To achieve the robust compute output for a column, we adopt a partial word-line activation (PWA) asserting only 8 rows rather than the entire 64 rows. This becomes necessary as low tunneling magnetoresistance (TMR) of MTJs involved in spin-based designs leads to severe adverse effects of the hardware non-idealities (such as the parasitic resistances associated with driver and sink circuitry [23]). Asserting more WLs leads to the larger $I_{BL}$ which increases the voltage drop across the parasitic resistances (also known as the loading effect) and aggravates the non-ideal effects. This leads to the deviation of the actual $I_{BL}$ from the ideal, increasing IMC errors [23]. Therefore, to enhance the sensing accuracy by mitigating this deviation of $I_{BL}$ from the expected (ideal) value, only a subset of rows (in our case, 8 rows) are asserted in one cycle. Hence, for a complete MAC operation, $N_R/8$ cycles are needed (where $N_R$ is the number of rows in the array). It is noteworthy that PWA or other techniques to mitigate the effect of non-idealities are also needed for other spin-based designs [8] and such effect is primarily due to the low TMR of MTJs. Moreover, PWA is a common technique to mitigate the energy and area overheads of ADCs, not just in BNNs, but also in high precision neural networks [25].

To ensure sufficient robustness, we perform a detailed analysis of the SM for different outputs. Figure 5(b) shows the $I_{BL}$ of column 0 with respect to $a$, which ranges from 0 to 8 in our design. The remaining columns are programmed to have equal distributions of '+1' and '-1' weights. We obtain $I_{BL}$ for each output considering different cases to ensure sensing robustness for all *IN-W* combinations. For that, we perform two sweeps representing the extreme cases for each output. Recall, for each column, we compute the dot product for input and weight vectors that are of size 8 each. Let us define the number of +1s in the applied input vector as $N_{IP}$ and that in the weight vector as $N_{WP}$. Thus, the number of -1s in the input and weight vectors are 8-$N_{IP}$ and 8-$N_{WP}$, respectively. First, we apply the *IN* sweep to obtain different outputs. In this, *W* is fixed as '+1' for the 8 rows (i.e., $N_{WP}$= 8), and $N_{IP}$ is swept from 0 through 8. In other words, *IN* vector is swept from {-1, …, -1} ($N_{IP}$ = 0; $a$ = 0) through {+1, -1, …, -1} ($N_{IP}$ = 1; $a$ = 1) and so on, up till {+1, …, +1} ($N_{IP}$ = 8; $a$ = 8). The second sweep is *W* sweep, in which, *IN* is fixed as '+1' for all the 8 rows ($N_{IP}$= 8), and $N_{WP}$ is swept from 0 through 8. That is, *W* vectors of {-1, …, -1} ($N_{WP}$ = 0; $a$ = 0) through {+1, -1, …, -1} ($N_{WP}$ = 1; $a$ = 1) and so on, up till {+1, …, +1} ($N_{WP}$= 8; $a$ = 8) are stored to obtained different output. It is noteworthy that for the same $a$, *IN* and *W* sweeps are expected to have the same $I_{BL}$ defined in equation (1). This is true if the columns are isolated. However, due to the cross-point connections and the sneak current paths between various cells in the array, these two extreme sweeps may result in different $I_{BL}$ values. In other words, each $a$ is now represented by a range of $I_{BL}$ ($I_{BLMIN}$ to $I_{BLMAX}$) corresponding to the *IN* and *W* sweeps [26], [27]. Therefore, the SM for a given $a$ and $a$-1 outputs is defined as ($I_{BL,MIN,a}$ - $I_{BL,MAX,a-1}$)/2. Further, the loading effect leads to non-linear behavior of $I_{BL}$ with respect to $a$, which also affects SM. In our design, we minimize the loading effect by co-optimizing the XNOR-VSH device and array, as will be discussed in the next section. Our results in Fig. 5(b) show that our proposed XNOR-VSH array exhibits a high linearity between $I_{BL}$ and $a$, and sufficiently large SM (> 1 $\mu A$ [26]) is obtained. The SM will be compared to other MTJ-based alternatives and discussed shortly in the following section.

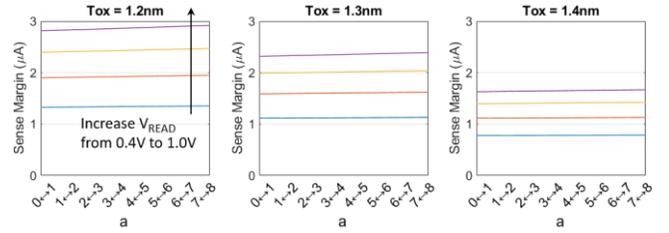

Fig. 6. Device-circuit co-optimization of sense margin (SM) in the proposed XNOR-VSH array with 8 rows asserted. The read voltage ($V_{READ}$) and MTJ tunneling oxide thickness ($T_{OX}$) are considered for SM optimization.

## IV. RESULTS AND ANALYSIS

In this section, we evaluate the proposed XNOR-VSH, and present its comparison with the existing MTJ-based XNOR-IMC designs. For the comparison, STT- and SOT-MRAMs are utilized to implement XNOR arrays with array size of 64*64. The co-optimization of device and array is performed for each design including the MTJ dimensions (see table I) and access transistor size. MTJs in STT- and VSH-based designs employ PMA while the MTJ in SOT-based design utilizes IMA. Each FL is designed to have the $E_B$ of ~50. The TMR of the MTJs is obtained as ~500 % for all designs, similar to the study in [23] and as experimentally obtained in [28]. The number of fins (*nfin*) of access transistors in STT- and SOT-based bit-cells are determined by considering multiple factors such as writability and energy-efficiency (details discussed in section IV. C).

### A. IMC Robustness

We follow the SM analysis methodology presented in Section III. B for the proposed XNOR-VSH array. Along with our design, the compute of the STT- and SOT-based XNOR arrays are also investigated using the same method and utilizing PWA (with 8 rows asserted) for comparison. As all three designs are based on the analog current summation, it is important to ensure by design that there is sufficient SM for compute accuracy. The SM is calculated for all values of $a$ that ranges from 0 to 8 (e.g. between 0 ↔ 1, …, 7 ↔ 8; see Fig. 5(b)). The worst-case SM of STT, SOT, and VSH-based arrays are 3.93 $\mu A$, 3.94 $\mu A$, and 1.12 $\mu A$, respectively. The lower SM in the proposed VSH-based design is owing to fact that the read path is formed along the WSe$_2$ channel layer, which is less conductive due to the low mobility of $WSe_2$ as well as Schottky contact resistance. On the other hand, STT- and SOT-based bit-cells have highly conductive FinFETs as access transistors in the read path. However, despite this disadvantage, we ensure that the proposed XNOR-VSH exhibits SM > 1 $\mu$A by performing a device-array co-optimization with the MTJ tunneling oxide (MgO) thickness and $V_{READ}$ optimization. The SM results in Fig. 6 show that the larger $V_{READ}$ and thinner MgO boost the SM. However, lowering the MgO thickness also leads to cell TMR degradation and aggravated effects of hardware non-idealities (as the sensing/driver resistance start becoming more dominant) [23]. Further, larger $V_{READ}$ or thinner MgO leads to an increase in the IMC energy and the read disturbance during IMC (discussed shortly). Considering all these aspects, we choose $V_{READ}$ of 0.4 V and MgO thickness of 1.3 nm as the optimal design point for our XNOR-VSH. Similarly, for XNOR-STT/-SOT, $V_{READ}$ of 0.1/0.2 V and MgO thickness of 1.1/1.3 nm are chosen from the optimization.

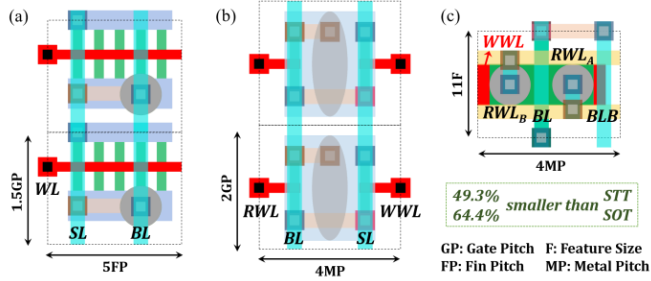

Fig. 7. Layout comparison of XNOR cells based on (a) STT-MRAM (5 fins), (b) SOT-MRAM (1 fin), and (c) VSH-MRAM. Note, STT- and SOT-based designs require two bit-cells to achieve XNOR operation.

Let us explain the MTJ disturb issue in a bit more detail. As currents flow along the MTJs ($I_{MTJ}$) during read/IMC, all the designs are under influence of $I_{MTJ}$-driven STT, which can disturb the MTJ states. Such a disturbance needs to be studied to ensure that the weights are retained during IMC. In this work, we quantify this using read disturb margin (RDM), which is defined as ($I_{CR} - I_{MTJ}$)/$I_{CR}$*100 (higher, the better). Here, $I_{CR}$ is the critical current associated with STT, above which the MTJ state is disturbed and an undesired switching happens. Given the $I_{MTJ}$ direction, XNOR-STT is vulnerable to AP to P switching with $I_{CR,AP \to P}$ ~ 14.2 $\mu$A. For XNOR-SOT, P to AP switching can occur with $I_{CR,P \to AP}$ = 54.7 $\mu$A. Due to the larger FL dimension, XNOR-SOT exhibits higher $I_{CR}$. The proposed XNOR-VSH array can be disrupted having both AP to P and P to AP switching as the MTJ currents flow in two directions (see Fig. 3 for details). The corresponding critical currents are $I_{CR,AP \to P}$ = 14.2 $\mu$A and $I_{CR,P \to AP}$ = 15.5 $\mu$A. Based on the currents obtained during IMC, we calculate the RDM of STT, SOT, and VSH-based XNOR-arrays as 88.1 %, 80.5 %, and 91.5 %, respectively. Note, the resistive $WSe_2$ channel in XNOR-VSH reduces the $I_{MTJ}$. This, while degrading SM (as discussed before) lowers the read disturb, because of which XNOR-VSH achieves the highest RDM among the three designs.

### B. Layout and Area Analysis

As mentioned before, the STT- and SOT-based XNOR cells utilize two bit-cells for complementary WE, limiting their scalability. However, in the proposed XNOR-VSH design, such an area penalty is avoided since a differential MTJ encoding in a single bit-cell is possible via VSH effect (see Fig. 7 and section III. A for details). Moreover, the integrated back-gate in the VSH device precludes the need for an additional access transistor. In this sub-section, we analyze the layout of each XNOR cell that consists of two STT/SOT bit-cells or one VSH bit-cell, and compare their area. Low area of XNOR-VSH is not only beneficial for high integration density but also leads other benefits such as IMC operation energy and latency, which is discussed in the following sub-section.

The layout analysis is performed using $\lambda$-based design rules [29], where $\lambda$ is half the minimum feature size (*F*) associated with technology. Gate pitch (GP), fin pitch (FP), and metal pitch (MP) are based on Intel 7nm technology [30]. As the STT- and SOT-based XNOR cells require two bit-cells (placed in a column) for XNOR-functionality, their cell heights are twice the height of the individual bit-cells. From the layout, we obtain area of each XNOR cell as 0.022 $\mu m^2$, 0.031 $\mu m^2$, and 0.011 $\mu m^2$ for STT, SOT, and VSH-based cells, respectively. Among the XNOR bit-cells, SOT-based design shows the largest area. The proposed XNOR-VSH exhibits 64.4 % lower area than XNOR-SOT and 49.3 % lower area than XNOR-STT.

### C. Latency and Energy Analysis

In this sub-section, we present the energy and latency analysis of the three designs considering write, read, and IMC. For comparison, STT, SOT, and VSH-based XNOR arrays (array size = 64*64) are designed with respective device-array co-optimizations as follows. In XNOR-STT array, the writability is ensured from using 5 fins in the access transistors in XNOR bit-cells and 1.1-nm-thick MgO in MTJs to flow sufficient write current. On the other hand, in XNOR-SOT array, ample write current along the HM is achieved even with 1 fin in the access transistors and 1.3-nm-thick MgO. A 1.3-nm-thick MgO is also used in XNOR-VSH, which is possible due to the decoupling of read and write paths. To maximize the immunity to read disturbance (discussed earlier) and reduce the effect of the load (sink/driver resistance), MgO thickness of 1.3 nm is effective in the XNOR-SOT and XNOR-VSH designs. From such device-array co-optimizations, we observe compactness-driven energy-efficiency (see Fig. 8) of the proposed XNOR-VSH array in all operations as discussed next.

#### 1) Write

The spin-based designs perform write (or WE) by switching the magnetization of the FLs of MTJs, which is controlled by the magnitude and *direction* of $I_{WRITE}$. Note, $I_{WRITE}$ flows through the MTJ in XNOR-STT, HM in XNOR-SOT, and $WSe_2$ channel in XNOR-VSH. In XNOR-STT and XNOR-SOT, $I_{WRITE}$ can flow only in one direction in a given cycle (since the BLs as well as the two bit-cells encoding differential weights are shared along the column). Thus, the FL is switched towards a specific direction in a cycle. Hence, to store the complementary weights for XNOR operation, two-cycles are needed - one to write P and the other to write AP - in XNOR-STT and XNOR-SOT. This increases the write latency and write power. On the other hand, the proposed XNOR-VSH can achieve the differential WE in a single bit-cell by the virtue of the VSH effect. Further, the inherent separation of spin currents in the two arms of the proposed device leads to differential WE in a *single* cycle (detailed in section III). To add to this, the VSH-based design inherently benefits write energy-efficiency due to (1) PMA-based write, (2) integrated back-gate-driven access-transistor-less compact layout (leading to lower BL/WL capacitances) and (3) large $\theta_{SH}$. As a result, we observe a significantly larger write efficiency in the proposed XNOR-VSH compared to the XNOR-STT and SOT. We achieve 72.2 % and 41.1 % lower write time, and 95.6% and 96.6% lower write energy in the XNOR-VSH array compared to XNOR-STT and XNOR-SOT arrays, respectively.

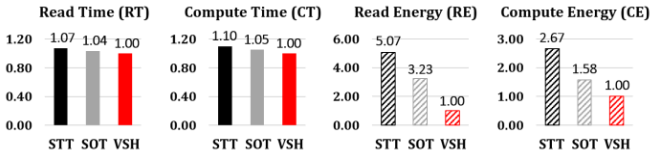

Fig. 8. Comparison of read/IMC energy and latency among STT, SOT, and VSH-based XNOR arrays (array size = 64*64).

*2) Read*

To analyze the read latency and energy, we assert one XNOR cell with IE corresponding to *IN* = +1 (as discussed earlier). The results show that compared to XNOR-STT and XNOR-SOT arrays, read time (RT) is lowered by 6.6 % and 3.4 %, and read energy (RE) is reduced by 80.3 % and 69.1 %, respectively. To explain the results, let us first discuss the differences in the read mechanisms of XNOR-STT/SOT and XNOR-VSH. Both XNOR-STT and XNOR-SOT perform the read operation by applying $V_{READ}$ (0.1 V for XNOR-STT and 0.2 V for XNOR-SOT) between BL and SL (see Fig. 1), following the conventional approach. Therefore, all 64 XNOR cells along the column contribute to the capacitances including wire capacitance and drain capacitance of access transistors. These capacitances need to be charged/discharged during the read operation, which entails energy consumption and latency. On the other hand, in XNOR-VSH, this overhead is largely mitigated. This is because the initial voltage values of all the lines are $V_{DD}$ and the read operation requires the application of $V_{DD}-V_{READ}$ to $RWL_A$ and 0 on WL, both of which are routed along the row. The BLs do not need to be charged/discharged during read (except for a small change in the BL voltage due to loading effect). Now, since multiple reads occur in parallel along different columns in an array, the energy cost associated with capacitive charging/discharging of $RWL_A$ and WL is amortized, leading to high read energy efficiency. Thus, the read energy savings are due to the compactness of the XNOR-VSH (leading to lower array capacitances) and the biasing scheme, which alleviate the BL energy consumption. The latency benefits of XNOR-VSH are mainly due to its compactness.

*3) In-Memory Compute (IMC)*

For IMC, we apply PWA in all three array designs, simultaneously activating 8 rows and 64 columns. The compute time (CT) and energy (CE) are calculated for the asserted 8 rows from one column (column 0). Note, other 63 columns (columns 1:63) affect the sensing current in column 0 during compute by modifying the equivalent resistance of the entire array but do not add CE. Among the three, the lowest CT and CE are observed in XNOR-VSH design: CT is 9.0 % and 4.8 % lower, and CE is 62.5 % and 36.6 % lower compared to STT- and SOT-based designs, respectively. These benefits stem from the factors similar to those discussed for read.

However, it is important to note some differences between IMC and read. In IMC, all designs consume larger WL charging/discharging energy compared to read due to the 8-row assertion. Moreover, higher sensing currents (i.e., accumulated $I_{BL}$) also play a role in increasing loading effect (which discharges BL to a lower value during IMC than read). This, in turn, increases the associated charging energy. Now, recall, XNOR-VSH achieves read benefits due to much lower BL charging component compared to other designs and amortization of $RWL_A$ and WL charging energies. In IMC, both these factors play a role but are less effective (due to high loading effect and multi-WL assertion), which reduces the IMC benefits of XNOR-VSH compared to read. However, the compactness of the proposed design still remains to be a major advantage, leading to the overall energy and latency benefits of the proposed design for IMC.

*D. Discussions and Future Outlook*

Our results establish the compactness and energy efficiency of XNOR-VSH vis-à-vis other spin-based memories (that need 2-bit-cells to encode signed weights). As noted earlier, the other approach of XNOR-IMC is to utilize a single bit-cell for unsigned binary IMC (AND-based IMC) and perform transformations between signed and unsigned regimes. This typically requires computing the average of the inputs and weights [12] and post-processing of the outputs, which incur some storage and compute costs. The proposed XNOR-VSH averts this penalty by performing in situ XNOR-IMC within the array. In this paper, we limit the quantitative comparisons to only the first XNOR-IMC approach (that performs IMC in the signed binary regime). The reason is that a proper qualitative comparison with the second XNOR-IMC would need to consider the fact that the VSH effect can also be utilized to design optimized AND-based IMC arrays, which along with transformations between signed and unsigned binary regimes can achieve XNOR-IMC. However, this requires an extensive optimization of VSH-based AND-IMC arrays, which is beyond the scope of this work. It may, however, be noted that the novelty of the approach in this paper lies in harnessing the VSH effect to avoid the transformations between signed and unsigned binary regimes, while encoding the differential weights in a single bit-cell.

Another point to note here is that we limit our analysis to comparing our technique with optimized solutions based on spin-memories only. As mentioned earlier, other XNOR-IMC solutions based on SRAM [6] and RRAM [7] exist. Our technique offers more compact design compared to SRAM and better endurance than RRAM, albeit with lower distinguishability. Therefore, simple benchmarking of the proposed technique with other technologies is challenging, as the role of endurance in BNNs needs to be accounted for a fair comparison, which is a subject of a future work.

## V. CONCLUSION

We propose a compact and energy-efficient array capable of performing in situ XNOR-based in-memory computation for BNNs. Our design, named XNOR-VSH, utilizes VSH effect in monolayer $WSe_2$. By the virtue of VSH, the proposed cell can perform XNOR-IMC with only one bit-cell compared to the state-of-the-art spin-based designs where, in general, two bit-cells are required for encoding complementary weights. The integrated back-gate in the proposed design further reduces the area by enabling access-transistor-less layout. These lead to a significant reduction in area compared to STT- and SOT-MRAM based XNOR designs. These benefits come at the cost of lower sensing robustness; but, by co-optimizing the XNOR-VSH device and array, we achieve sufficiently large sense margin ($> 1~\mu A$) for robust IMC. The compactness of XNOR-VSH along with other technological and array-level attributes offer improvements in the energy efficiency for write, read and IMC, thus leading to an energy-efficient and compact solution of BNNs.

ACKNOWLEDGMENT

The authors thank Akul Malhotra (Purdue University, West Lafayette, IN, USA) for his input on trained BNNs with the ResNet 18 architecture.

REFERENCES

[1] C. J. Jhang, C. X. Xue, J. M. Hung, F. C. Chang, and M. F. Chang, "Challenges and trends of SRAM-Based computing-in-memory for AI edge devices," *IEEE Trans. Circuits Syst. I Regul. Pap.*, vol. 68, no. 5, pp. 1773–1786, 2021, doi: 10.1109/TCSI.2021.3064189.

[2] W. Haensch, T. Gokmen, and R. Puri, "The Next Generation of Deep Learning Hardware: Analog Computing," *Proc. IEEE*, vol. 107, no. 1, pp. 108–122, 2019, doi: 10.1109/JPROC.2018.2871057.

[3] I. Chakraborty et al., "Resistive Crossbars as Approximate Hardware Building Blocks for Machine Learning: Opportunities and Challenges," *Proc. IEEE*, vol. 108, no. 12, pp. 2276–2310, 2020, doi: 10.1109/JPROC.2020.3003007.

[4] J.-M. Hung, C.-J. Jhang, P.-C. Wu, Y.-C. Chiu, and M.-F. Chang, "Challenges and Trends of Nonvolatile In-Memory-Computation Circuits for AI Edge Devices," *IEEE Open J. Solid-State Circuits Soc.*, vol. 1, no. September, pp. 171–183, 2021, doi: 10.1109/ojsscs.2021.3123287.

[5] M. Rastegari, V. Ordonez, J. Redmon, and A. Farhadi, "XNOR-Net: ImageNet classification using binary convolutional neural networks," in Proc. Eur. Conf. Comput. Vis. (ECCV), 2016.

[6] S. Yin, Z. Jiang, J. S. Seo, and M. Seok, "XNOR-SRAM: In-Memory Computing SRAM Macro for Binary/Ternary Deep Neural Networks," *IEEE J. Solid-State Circuits*, vol. 55, no. 6, pp. 1733–1743, 2020, doi: 10.1109/JSSC.2019.2963616.

[7] X. Sun, S. Yin, X. Peng, R. Liu, J. S. Seo, and S. Yu, "XNOR-RRAM: A scalable and parallel resistive synaptic architecture for binary neural networks," *Proc. 2018 Des. Autom. Test Eur. Conf. Exhib. DATE 2018*, vol. 2018-Janua, pp. 1423–1428, 2018, doi: 10.23919/DATE.2018.8342235.

[8] T. N. Pham, Q. K. Trinh, I. J. Chang, and M. Alioto, "STT-BNN: A Novel STT-MRAM In-Memory Computing Macro for Binary Neural Networks," *IEEE J. Emerg. Sel. Top. Circuits Syst.*, vol. 12, no. 2, pp. 569–579, 2022, doi: 10.1109/JETCAS.2022.3169759.

[9] S. Jung et al., "A crossbar array of magnetoresistive memory devices for in-memory computing," *Nature*, vol. 601, no. 7892, pp. 211–216, 2022, doi: 10.1038/s41586-021-04196-6.

[10] H. Wang, W. Kang, B. Pan, H. Zhang, E. Deng, and W. Zhao, "Spintronic Computing-in-Memory Architecture Based on Voltage-Controlled Spin-Orbit Torque Devices for Binary Neural Networks," *IEEE Trans. Electron Devices*, vol. 68, no. 10, pp. 4944–4950, 2021, doi: 10.1109/TED.2021.3102896.

[11] H. Kim, J. Sim, Y. Choi, and L. S. Kim, "NAND-net: Minimizing computational complexity of in-memory processing for binary neural networks," Proc. - 25th IEEE Int. Symp. High Perform. Comput. Archit. HPCA 2019, pp. 661–673, 2019, doi: 10.1109/HPCA.2019.00017.

[12] H. Oh, H. Kim, N. Kang, Y. Kim, J. Park, and J. J. Kim, "Single RRAM Cell-based In-Memory Accelerator Architecture for Binary Neural Networks," 2021 IEEE 3rd Int. Conf. Artif. Intell. Circuits Syst. AICAS 2021, pp. 2021–2024, 2021, doi: 10.1109/AICAS51828.2021.9458444.

[13] H. Kim, H. Oh, and J. J. Kim, "Energy-efficient XNOR-free In-Memory BNN Accelerator with Input Distribution Regularization," IEEE/ACM Int. Conf. Comput. Des. Dig. Tech. Pap. ICCAD, vol. 2020-Novem, 2020, doi: 10.1145/3400302.3415641.

[14] S. K. Thirumala et al., "Valley-Coupled-Spintronic Non-Volatile Memories With Compute-In-Memory Support," vol. 19, pp. 635–647, 2020.

[15] K. F. Mak, K. L. McGill, J. Park, and P. L. McEuen, "The valley hall effect in MoS2 transistors," Science (80-. )., vol. 344, no. 6191, pp. 1489–1492, 2014, doi: 10.1126/science.1250140.

[16] T. Y. T. Hung, K. Y. Camsari, S. Zhang, P. Upadhyaya, and Z. Chen, "Direct observation of valley-coupled topological current in MoS 2," Sci. Adv., vol. 5, no. 4, pp. 1–7, 2019, doi: 10.1126/sciadv.aau6478.

[17] T. Y. T. Hung, A. Rustagi, S. Zhang, P. Upadhyaya, and Z. Chen, " Experimental observation of coupled valley and spin Hall effect in p-doped WSe 2 devices ," InfoMat, vol. 2, no. 5, pp. 968–974, 2020, doi: 10.1002/inf2.12095.

[18] K. Cho, X. Liu, Z. Chen, and S. Gupta, "Utilizing Valley – Spin Hall Effect in Monolayer WSe 2 for Designing Low Power Nonvolatile Spintronic Devices and Flip-Flops," IEEE Trans. Electron Devices, vol. 69, no. 4, pp. 1667–1676, 2022.

[19] C. D. English, G. Shine, V. E. Dorgan, K. C. Saraswat, and E. Pop, "Improved contacts to MoS2 transistors by ultra-high vacuum metal deposition," Nano Lett., vol. 16, no. 6, pp. 3824–3830, 2016, doi: 10.1021/acs.nanolett.6b01309.

[20] K. Y. Camsari, S. Ganguly, and S. Datta, "Modular approach to spintronics," Sci. Rep., vol. 5, pp. 1–13, 2015, doi: 10.1038/srep10571.

[21] X. Fong, S. K. Gupta, N. N. Mojumder, S. H. Choday, C. Augustine, and K. Roy, "KNACK: A hybrid spin-charge mixed-mode simulator for evaluating different genres of spin-transfer torque MRAM bit-cells," Int. Conf. Simul. Semicond. Process. Devices, SISPAD, pp. 51–54, 2011, doi: 10.1109/SISPAD.2011.6035047.

[22] C. F. Pai, L. Liu, Y. Li, H. W. Tseng, D. C. Ralph, and R. A. Buhrman, "Spin transfer torque devices utilizing the giant spin Hall effect of tungsten," Appl. Phys. Lett., vol. 101, no. 12, pp. 1–5, 2012, doi: 10.1063/1.4753947.

[23] T. Sharma, C. Wang, A. Agrawal, and K. Roy, "Enabling Robust SOT-MTJ Crossbars for Machine Learning using Sparsity-Aware Device-Circuit Co-design," Proc. Int. Symp. Low Power Electron. Des., vol. 2021-July, 2021, doi: 10.1109/ISLPED52811.2021.9502492.

[24] S. Jain, S. K. Gupta, and A. Raghunathan, "TiM-DNN: Ternary In-Memory Accelerator for Deep Neural Networks," IEEE Trans. Very Large Scale Integr. Syst., vol. 28, no. 7, pp. 1567–1577, 2020, doi: 10.1109/TVLSI.2020.2993045.

[25] K. He, I. Chakraborty, C. Wang, and K. Roy, "Design space and memory technology co-exploration for in-memory computing based machine learning accelerators," IEEE/ACM Int. Conf. Comput. Des. Dig. Tech. Pap. ICCAD, 2022, doi: 10.1145/3508352.3549453.

[26] N. Thakuria, R. Elangovan, S. K. Thirumala, A. Raghunathan, and S. K. Gupta, "STeP-CiM: Strain-Enabled Ternary Precision Computation-In-Memory Based on Non-Volatile 2D Piezoelectric Transistors," Front. Nanotechnol., vol. 4, pp. 1–6, 2022, doi: 10.3389/fnano.2022.905407.

[27] K. Cho, X. Fong, and S. K. Gupta, "Exchange-Coupling-Enabled Electrical-Isolation of Compute and Programming Paths in Valley-Spin Hall Effect based Spintronic Device for Neuromorphic Applications," Device Res. Conf. - Conf. Dig. DRC, vol. 2021-June, no. 70, pp. 7–8, 2021, doi: 10.1109/DRC52342.2021.9467139.

[28] S. Ikeda et al., "Tunnel magnetoresistance of 604% at 300 K by suppression of Ta diffusion in CoFeBMgOCoFeB pseudo-spin-valves annealed at high temperature," Appl. Phys. Lett., vol. 93, no. 8, 2008, doi: 10.1063/1.2976435.

[29] [Online]. Available: http://www.mosis.com/pages/design/rules/index (May 2009).

[30] [Online]. Available: https://en.wikipedia.org/wiki/7_nm_process.

[31] S. Ikeda et al., ``A perpendicular-anisotropy CoFeB-MgO magnetic tunnel junction,'' Nature Mater., vol. 9, no. 9, pp. 721-724, 2010, doi: 10.1038/nmat2804.

[32] R. Andrawis, A. Jaiswal, and K. Roy, ``Design and comparative analysis of spintronic memories based on current and voltage driven switching,'' IEEE Trans. Electron Devices, vol. 65, no. 7, pp. 2682-2693, Jul. 2018.